\documentstyle[prl,aps]{revtex}
\input BoxedEPS.tex
\SetOzTeXEPSFSpecial
\HideDisplacementBoxes

\begin{document}

%\draft{}
\twocolumn[\hsize\textwidth\columnwidth\hsize
          \csname @twocolumnfalse\endcsname
\title{Reply to ``Comment on `Evidence for the immobile bipolaron 
formation in the paramagnetic state of the magnetoresistive manganites'''}
\author{Guo-meng Zhao} 
\address{Department of Physics and Astronomy, California State University at Los Angeles, 
Los Angeles, CA 90032, USA}

\maketitle
\widetext

\begin{abstract}

Zhao {\em et al.} (Phys. Rev. B {\bf 62}, R11 949 (2000)) reported studies of the oxygen-isotope effects on the 
intrinsic resistivity and thermoelectric power in the paramagnetic 
state of several 
ferromagnetic manganites. The isotope effects on the intrinsic 
electrical transport properties are not 
consistent with a simple small-polaron hopping mechanism, but can be 
well explained by a model where there coexist immobile bipolarons and 
thermally-excited small polarons.  Recently, Banerjee {\em et al.} 
(Phys.  Rev.  B {\bf 68}, 186401 (2003)) wrote a Comment on our paper, 
where they misunderstood our theoretical model, erroneously thinking 
that we explain the electrical transport in the paramagnetic state as 
due to bipolaron hopping.  They confuse large polarons with small 
bipolarons, use incorrect conditions for bipolaron formation, and even 
incorrectly reproduce our data.  We show that the model used in the 
Comment is inconsistent with any features of the observed 
oxygen-isotope effects.  More evidence is provided to support the 
theoretical model used in our original paper.
\end{abstract}
\vspace{0.5cm}
\narrowtext
%\newpage
]
Experimental evidence for small polaron charge carriers in the 
paramagnetic state was provided by transport measurements \cite{Jaime}.  
It was found that the activation energy $E_{\rho}$ deduced from the 
conductivity data is one order of magnitude larger than the activation 
energy $E_{s}$ obtained from the thermoelectric power data.  Such a 
large difference in the activation energies is the hallmark of the 
small-polaron hopping conduction \cite{Jaime}.  In our recent paper 
\cite{ZhaoPM}, we reported studies of the oxygen-isotope effects on the 
intrinsic resistivity and thermoelectric power in the paramagnetic 
state of several ferromagnetic manganites.  The isotope effects on the 
intrinsic electrical transport properties can be well explained by a 
model where there coexist immobile bipolarons and thermally-excited 
small polarons.  In our theoretical model, the electrical transport in 
the paramagnetic state is due to small polaron hoping, but such 
polarons are thermally excited from localized bipolaronic states.  
Recently, Banerjee {\em et al.} \cite{Ban} wrote a Comment on our 
paper, where they misunderstood our theoretical model, erroneously 
thinking that we explain the electrical transport in the paramagnetic 
state as due to bipolaron hopping.  They confuse large polarons with 
small bipolarons, use incorrect conditions for bipolaron formation, 
and even incorrectly reproduce our data.  We show that the model used 
in the Comment is inconsistent with any features of the observed 
oxygen-isotope effects.  More evidence is provided to support the 
theoretical model used in our original paper \cite{ZhaoPM}.

Fig.~1 shows the temperature dependence of the resistivity of the 
oxygen-isotope exchanged films of La$_{0.75}$Ca$_{0.25}$MnO$_{3}$ and  
Nd$_{0.7}$Sr$_{0.3}$MnO$_{3}$.  The data are the same as those reported in 
our original paper \cite{ZhaoPM}. The only difference is that the 
resistivity is plotted as a function of $T/T_{C}$ in our original 
paper \cite{ZhaoPM}, where $T_{C}$ is the Curie temperature.  Comparing the data shown here and 
the data shown in Fig.~1 of the Comment \cite{Ban}, we can clearly see that they 
do not reproduce our data correctly.  Further, the authors of the 
Comment \cite{Ban} fit their incorrectly reproduced data by
\begin{equation}\label{COM}
\rho = DT\exp(E_{\rho}/k_{B}T),
\end{equation}
where the coefficient $D$ is inversely proportional 
to the characteristic phonon frequency $\nu_{0}$.

\begin{figure}[htb]
    \ForceWidth{6.5cm}
\centerline{\BoxedEPSF{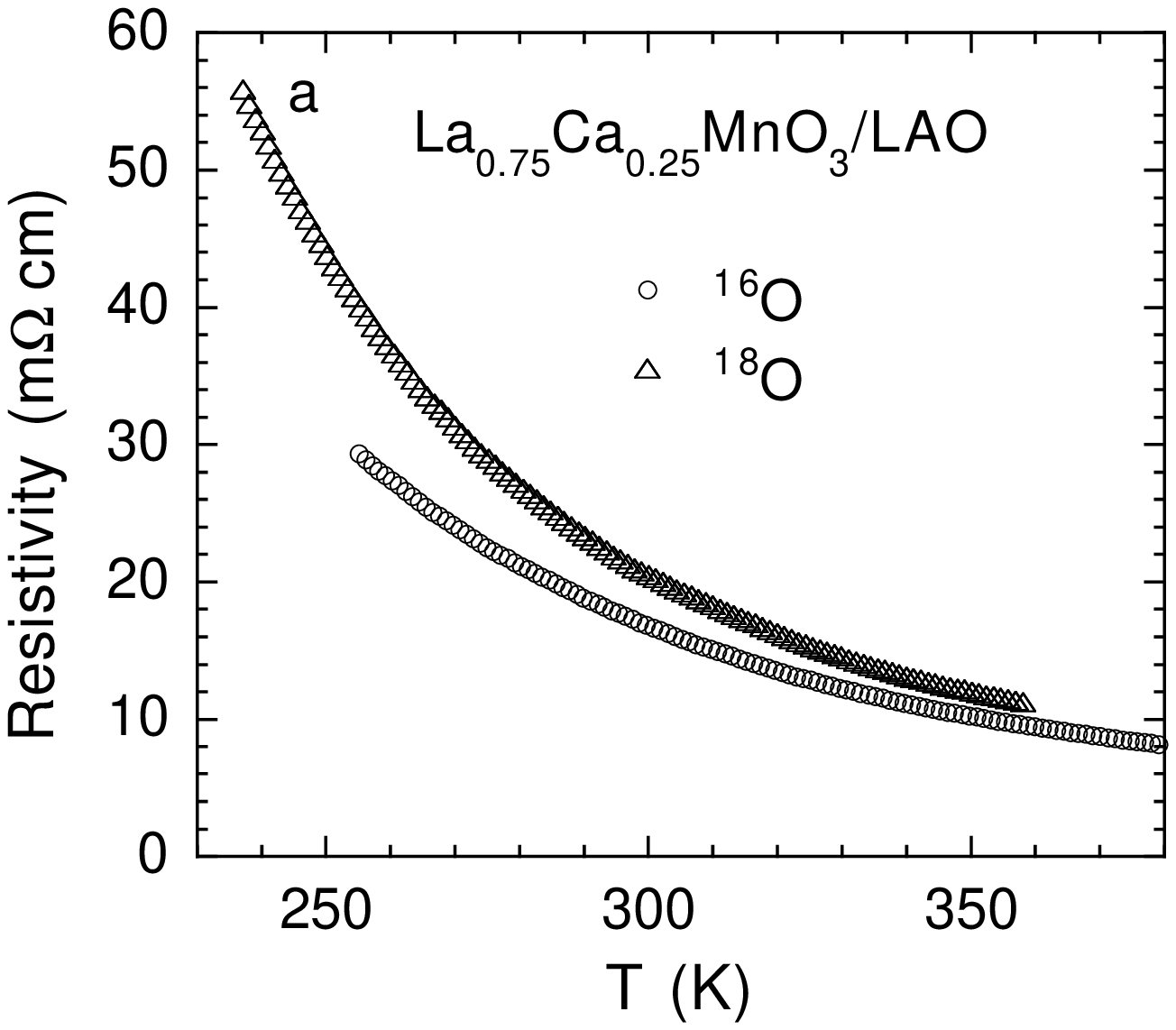}}
\ForceWidth{7cm}
\centerline{\BoxedEPSF{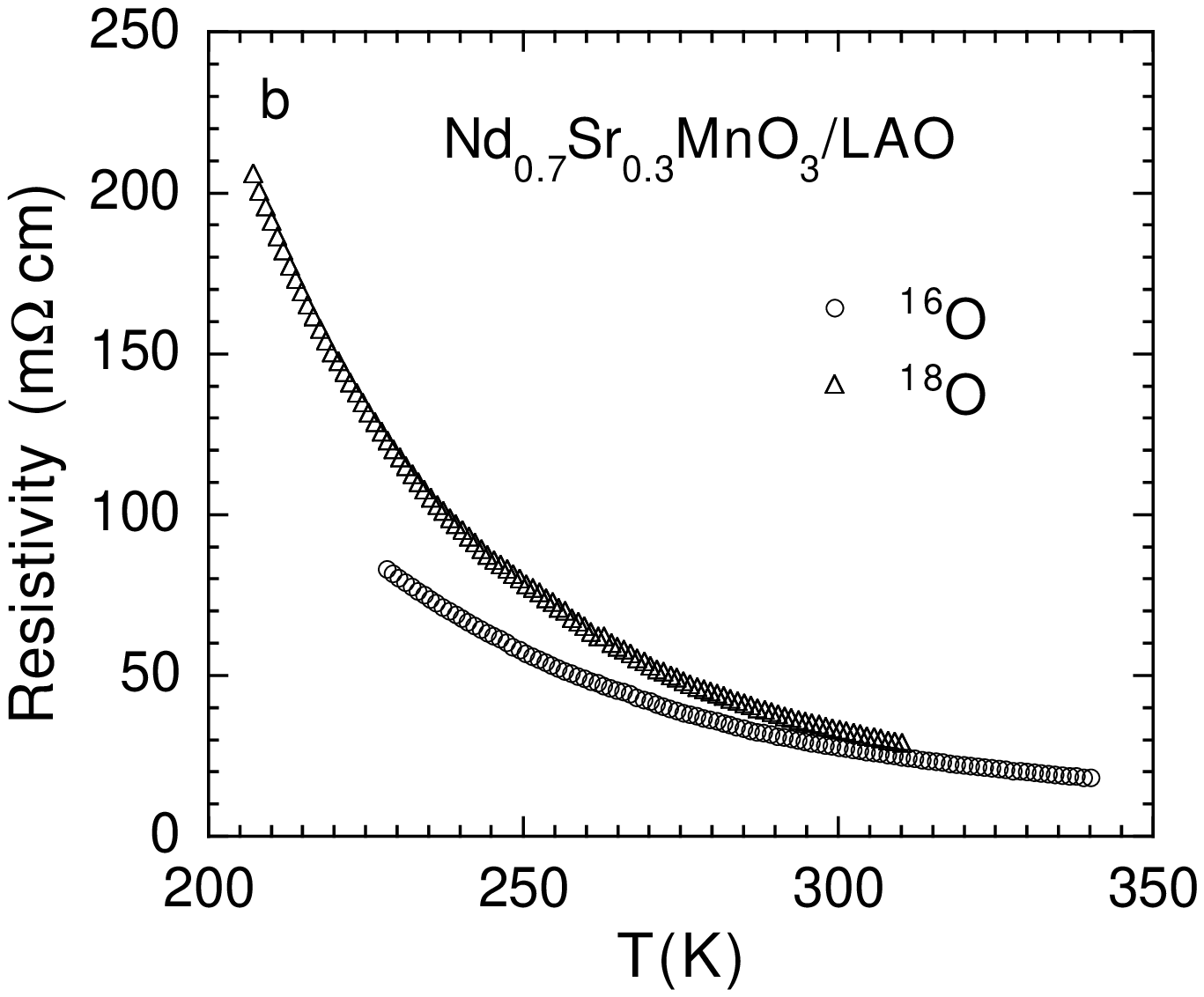}}
	\vspace{0.3cm}
\caption[~]{The temperature dependence of the resistivity of the 
oxygen-isotope exchanged films of a) La$_{0.75}$Ca$_{0.25}$MnO$_{3}$ and 
b) Nd$_{0.7}$Sr$_{0.3}$MnO$_{3}$.  The data are the same as those reported in 
Ref.~\cite{ZhaoPM}. }
	\protect\label{Fig.1}
\end{figure}

We can also fit the data of Fig.~1 by Eq.~\ref{COM}.  The  
curves fitted to the resistivity data of the oxygen-isotope exchanged La$_{0.75}$Ca$_{0.25}$MnO$_{3}$ films are shown 
in Fig.~2.  The best fits yield the fitting parameter $D$ = 7.36$\times$10$^{-7}$ $\Omega$cm/K 
for the $^{16}$O sample and $D$ = 6.15$\times$10$^{-7}$ $\Omega$cm/K for the 
$^{18}$O sample.  We can see that the $D$ value for the $^{16}$O 
sample is larger than that for the $^{18}$O sample by 20$\%$.  Since 
$D$ $\propto$
1/$\nu_{0}$ $\propto$ $\sqrt{M}$, we expect that the $D$ value for the 
$^{16}$O sample should be smaller than that for the $^{18}$O sample.  
This is in sharp contrast with the isotope effect on the $D$ value obtained 
above. This implies that Eq.~\ref{COM} is not a 
relevant formula to describe the electrical transport in the paramagnetic state 
of La$_{0.75}$Ca$_{0.25}$MnO$_{3}$.  If the authors of the Comment would 
provide the fitting parameters for both isotope samples, they would not 
claim that their model could better explain our data.

\begin{figure}[htb]
\ForceWidth{7cm}
\centerline{\BoxedEPSF{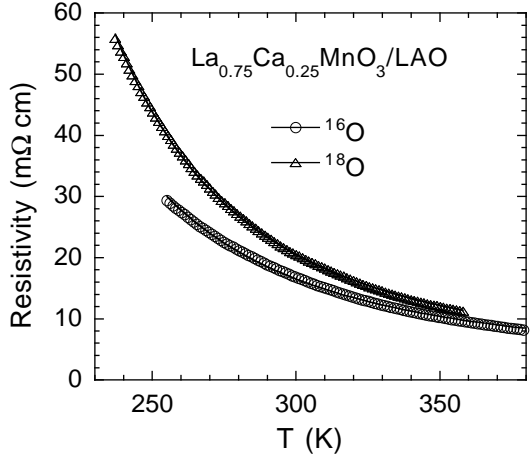}}
	\vspace{0.6cm}
\caption[~]{The temperature dependence of the resistivity of the 
oxygen-isotope exchanged films of La$_{0.75}$Ca$_{0.25}$MnO$_{3}$.  
The solid lines are the curves best fitted by Eq.~\ref{COM}.  The 
fitting parameter $D$ for the $^{16}$O sample is larger than that for the 
$^{18}$O sample by 20$\%$.  }
	\protect\label{Fig.2}
\end{figure}

In fact, we have shown \cite{ZhaoPM} that if $k_{B}T$ is less 
than the polaron bandwidth $W_{p}$ and only thermally excited polarons from 
localized bipolaron states contribute to the electrical transport in 
the paramagnetic state, the resistivity is given by
\begin{equation}\label{REP}
\rho = \frac{C}{\sqrt{T}}\exp(E_{\rho}/k_{B}T),
\end{equation}
where $E_{\rho} = E_{a} + E_{s}$, $C = 
(ah/e^{2}\sqrt{k_{B}})(1.05W_{p})^{1.5}/h\nu_{o}$, $a$ is the hopping 
distance,  $E_{s}$ is 
the activation energy for polarons to be thermally excited from 
localized bipolaron states, and $E_{a}$ depends on the polaron binding energy 
$E_{p}$ and bare hoping integral $t$. The quantity $C$ should strongly 
depend on the isotope mass $M$ and decrease with increasing $M$.  This 
is because $W_{p}$ decreases strongly with increasing $M$ according to 
$W_{p}$ = $12t\exp(-\Gamma E_{p}/\hbar\omega_{o})$ = $12t\exp(-g^{2})$, 
where $\Gamma$ is a constant ($<$1) \cite{Alex99,Alex99C}.  In our 
original paper \cite{ZhaoPM}, we fit the resistivity data of the 
$^{16}$O and $^{18}$O samples of La$_{0.75}$Ca$_{0.25}$MnO$_{3}$ by 
Eq.~\ref{REP} and find that the parameter $C$ for the $^{16}$O sample 
is larger than that for the $^{18}$O sample by 34$\%$, in agreement
with the
theoretical model
(Eq.~\ref{REP}).  In fact, this model can quantitatively explain \cite{ZhaoPM} 
the combined isotope effects on $C$ and on $E_{s}$.

On the other hand, if small polarons are bound to localized impurity 
states rather than to localized bipolaron states, $E_{s}$ would be 
independent of the isotope mass in the zero-order approximation 
according to Austin and Mott \cite{Austin}.  In our original paper, we 
have only considered this possibility \cite{ZhaoPM}.  Here, we shall 
consider the other possibility, that is, a polaron is bound to an 
impurity site with the screened Coulombic interaction reduced by a 
factor of the static dielectric constant $\epsilon_{s}$ (continuous 
medium approximation).  The binding energy of the impurity state is 
given by \cite{Book}
\begin{equation}\label{EC}
 E_{c} = \frac{13.6}{\epsilon_{s}^{2}}\frac{m^{*}}{m_{e}}~eV,
\end{equation}
where $m^{*}$ is 
the effective mass of a polaron and $m_{e}$ is the mass of a free 
electron. The thermally excited polaron density from the impurity 
states is \cite{Book}
\begin{equation}
n = (2N_{d})^{1/2}(m^{*}k_{B}T/2\pi\hbar^{2})^{3/4}\exp(- E_{c}/2k_{B}T),
\end{equation}
where $N_{d}$ is the dopant concentration. The thermoelectric 
power is correspondingly given by \cite{Austin}
\begin{equation}\label{Th}
S = \frac{k_{B}}{e} (E_{s}/k_{B}T + \alpha^{\prime}),
\end{equation}
where $E_{s}$ = $E_{c}/2$ and $\alpha^{\prime}$ is a constant depending 
on the kinetic energy of the polarons and on the polaron density 
\cite{Austin}. With the mobility of 
polarons \cite{Emin} $\mu$ $\propto$ $\nu_{o}/k_{B}T\exp (- E_{a}/k_{B}T)$, we 
finally have
\begin{equation}\label{REP1}
\rho = BT^{0.25}\exp(E_{\rho}/k_{B}T),
\end{equation}
where $E_{\rho} = E_{a} + E_{c}/2$ and $B$ $\propto$ $W_{p}^{3/4}/\nu_{o}$.

In Fig.~3 we fit the resistivity data of the isotope exchanged 
La$_{0.75}$Ca$_{0.25}$MnO$_{3}$ by Eq.~\ref{REP1}.  The best fits yield 
$B$ = 1.13$\times$10$^{-4}$ $\Omega$cm/K$^{0.25}$ 
for the $^{16}$O sample and $B$ = 0.89$\times$10$^{-4}$ 
$\Omega$cm/K$^{0.25}$ for the $^{18}$O sample.  It is clear that the 
$B$ value for the $^{16}$O sample is larger than that for the $^{18}$O 
sample by 27$\%$.  This is in agreement with the theoretical 
prediction.  If we assume that $\nu_{o}$ is inversely proportional to 
the square root of the reduced mass of the manganese and oxygen 
atoms, $\nu_{o}$ will decrease by 4.6$\%$ upon
replacing $^{16}$O by $^{18}$O.  Using the oxygen-isotope effect on 
$B$ and the relation $B$ $\propto$ $W_{p}^{3/4}/\nu_{o}$, we find that 
$W_{p}$ for the $^{18}$O sample is smaller than that for 
the $^{16}$O sample by 46$\%$.  The exponent of the oxygen-isotope effect on 
$W_{p}$ is $\beta_{O}$ = $-d\ln W_{p}/d\ln M_{O}$ = 3.2, where $M_{O}$ 
is the oxygen mass.

\begin{figure}[htb]
\ForceWidth{7cm}
\centerline{\BoxedEPSF{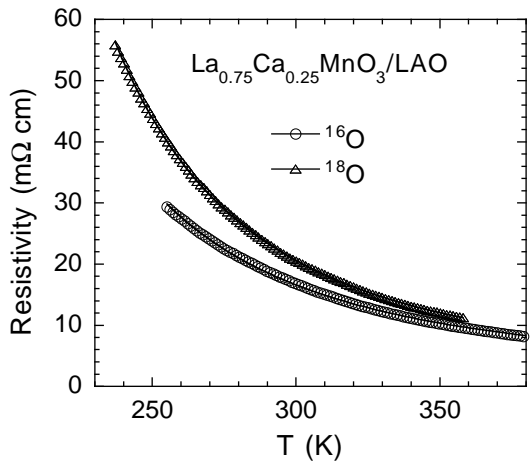}}
	\vspace{0.3cm}
\caption[~]{The temperature dependence of the resistivity of the 
oxygen-isotope exchanged films of La$_{0.75}$Ca$_{0.25}$MnO$_{3}$.  
The solid lines are the curves best fitted by Eq.~\ref{REP1}.  The 
fitting parameter $B$ for the $^{16}$O sample is larger than that for 
the $^{18}$O sample by 27$\%$.  }
	\protect\label{Fig.3}
\end{figure}

From the thermoelectric power data of the isotope exchanged 
La$_{0.75}$Ca$_{0.25}$MnO$_{3}$, we have found that \cite{ZhaoPM} $E_{s}$ = 
13.2 meV for the $^{16}$O sample and $E_{s}$ = 18.7 meV for the 
$^{18}$O sample.  Then we have $E_{c}$ = 26.4 meV for the $^{16}$O 
sample and $E_{c}$ = 37.4 meV for the $^{18}$O sample.  Using 
Eq.~\ref{EC} and $W_{p}$ $\propto$ $1/m^{*}$, we find $\beta_{O}$ = 
3.0, close to the value (3.2) deduced from the resistivity data above.  This 
appears to indicate that this model can quantitatively explain the 
isotope effects.

From $W_{p}$ = $12t\exp(-\Gamma E_{p}/\hbar\omega_{o})$ = 
$12t\exp(-g^{2})$, we can easily show that $\beta_{O}$ = $g^{2}/2.6$.  
Here we have used the fact that $\omega_{o}$ is inversely proportional to 
the square root of the reduced mass of the manganese and oxygen 
atoms.  Using $\beta_{O}$ = 3.0-3.2, one has $g^{2}$ = 7.8-8.3.  Assuming 
that the bare effective mass of charged carriers is about 0.5$m_{e}$, 
we find that the effective mass of polarons $m^{*}$ is about 
1500$m_{e}$, which seems unlikely.

It is worthy of noting that Eq.~\ref{EC} is valid only if the Bohr 
radius $r_{B}$ = $0.53\epsilon_{s} m_{e}/m^{*}$ \AA~is significantly 
larger than the interatom distance (2 \AA) \cite{Book}.  Using 
$\epsilon_{s}$ = 20 (Ref.~\cite{Kat}) and $m^{*}$ = 1500$m_{e}$, we 
get $r_{B}$ = 0.007 \AA~and $E_{c}$ = 51 eV.  The calculated $E_{c}$ = 
51 eV is over three orders of magnitude larger than the measured one.  
The calculated $r_{B}$ = 0.007 \AA~implies that the continuous medium 
approximation does not hold. Thus Eq.~\ref{EC} cannot be applied to doped manganites 
unless the effective mass of charged carriers is close to $m_{e}$.  
This is possible only if the charged carriers are not small polarons, 
in contradiction with the large isotope exponent $\beta_{O}$ and the 
observed small polaron hopping conduction.  Therefore, the model where 
polarons are bound to impurity sites with the screened Coulombic 
interaction (continuous medium approximation) cannot explain the 
observed isotope effects on the intrinsic electrical transport properties.

On the other hand, if polarons are bound into much heavier 
bipolarons \cite{Alexcond}, the random potentials produced by dopants can completely 
localize bipolaron band.  The minimum separation between the polaron 
and bipolaron bands is $\Delta = 2(1-\Gamma)E_{p}- V_{c}- W_{p}$, 
where $V_{c}$ is the Coulombic repulsion between bound polarons 
\cite{Alex99C}.  It is clear that the value of $\Delta$ could be 
less than 0.1 eV, in agreement 
with the measured value of about 30 meV in 
La$_{0.75}$Ca$_{0.25}$MnO$_{3}$.

Within the same model where polarons are bound into the localized bipolaron 
states \cite{ZhaoPM}, we have also determined the bandwidths for the isotope exchanged 
samples of La$_{0.75}$Ca$_{0.25}$MnO$_{3}$, that is, $W_{p}$ = 49.0 
meV for $^{16}$O sample and $W_{p}$ = 38.8 meV for the $^{18}$O 
sample.  The values of $W_{p}$ lead to $\beta_{O}$ = 2.0 and $g^{2}$ = 
4.6.  Then the bare hopping integral $t$ is calculated to be 0.41 eV 
from the values of $g^{2}$ and $W_{p}$. The inferred value of $t$ is in 
reasonable agreement with the band structure calculation \cite{Pick1}.  
Further, since $E_{s}$ = 
$\Delta/2$ and $\Delta = 2(1-\Gamma)E_{p}- V_{c}- W_{p}$, the change 
of $E_{s}$ due to the oxygen isotope substitution is $\delta E_{s}$ = 
$\delta\Delta/2 = - \delta W_{p}/2$.  With $\delta W_{p}/2$ = - 11.2 
meV for La$_{0.75}$Ca$_{0.25}$MnO$_{3}$, one has $\delta E_{s}$ = 5.5 
meV, in quantitative agreement with the value (5.5$\pm$0.6 meV) 
deduced independently from the thermoelectric power data 
\cite{ZhaoPM}.

The electron-phonon coupling constant $g^{2}$ can be also determined 
from optical conductivity data.  It is shown \cite{Alex99C} that the optical 
conductivity contributed from polaronic charged carriers exhibits a 
broad peak feature at an energy of $E_{m}$ = 2$g^{2}\hbar\omega_{o}$.  
Jung {\em et al.} \cite{Jung} have identified the polaronic conductivity 
and determined the value of $E_{m}$ as a function of doping $x$ in 
La$_{1-x}$Ca$_{x}$MnO$_{3}$ system.  From their results \cite{Jung}, 
one can easily see that $E_{m}$ $\simeq$ 0.7 eV for $x$ = 0.25.  If we 
take $\hbar\omega_{o}$ = 0.074 eV (Ref.~\cite{ZhaoPM}), we have 
$g^{2}$ = 4.7, in quantitative agreement with the value (4.6) deduced 
from the isotope effects.

In our data analyses \cite{ZhaoPM}, we assume that bipolarons are completely 
localized by random potential due to dopants and defects, and 
the conductivity is only contributed from thermally excited small polarons.  
The authors of the Comment misunderstand our theoretical model, 
erroneously thinking that we explain the conductivity in the 
paramagnetic state as due to bipolaron hopping.  Furthermore, they 
confuse large polarons with small bipolarons.  From the value of 
$\alpha^{\prime}$ in Eq.~\ref{Th}, which is found to be less than 1 
for our samples, the authors of the Comment conclude that \cite{Ban} 
our data are consistent with small polaron hopping conduction 
mechanism, but disagree with bipolaron hopping conduction mechanism.  
This is incorrect.  The value of $\alpha^{\prime}$ can only 
discriminate between small polaron and large polaron conduction 
mechanisms.  Bipolarons are not equivalent to large polarons.

Finally, the arguments \cite{Ban} against the formation of bipolarons 
in doped manganites do not have scientific ground.  The authors of the 
Comment argue that bipolarons cannot be formed in doped manganites 
because there is not enough disorder.  We do not believe that disorder 
is a required condition for the formation of bipolarons.  Bipolarons 
can be formed when electron-phonon interactions are strong enough to 
overcome direct Coulombic interaction between two polarons 
\cite{Alex99C}.  The authors of the Comment \cite{Ban} also mistakenly 
think that the charge disproportionality is a necessary condition for 
the formation of bipolarons.  Intersite bipolarons do not involve any 
charge disproportionality.  In this Comment, there is also a mistake 
in the criterion for the adiabatic or nonadiabatic hoping 
conduction mechanisms. The bare bandwidth should be used to compare the 
parameter $\phi$ = 
$(k_{B}TE_{p}/\pi)^{1/4}(\hbar\omega_{o}/\pi)^{1/2}$.  But the authors 
of the Comment \cite{Ban} use the effective polaron bandwidth instead 
of the bare bandwidth to compare the parameter $\phi$.  It is clear 
that the bare bandwidth is larger than $\phi$ so that the adiabatic 
hopping conduction is relevant in doped manganites.
~\\
~\\
{\bf Acknowledgment:} This research was partly supported by an award from 
Research Corporation.
~\\
~\\
* Correspondence should be addressed to gzhao2@calstatela.edu.


\begin{thebibliography}{99}

\bibliographystyle{prsty}

\bibitem{Jaime} M.  Jaime, M.  B.  Salamon, 
M.  Rubinstein, R.  E.  Treece, J.  S.  Horwitz, and D.  B.  Chrisey, 
Phys.  Rev.  B \textbf{54}, 11914 (1996).

\bibitem{ZhaoPM}G. M. Zhao,Y.  S.  Wang, D.  J.  Kang, W.  Prellier, M.  
Rajeswari, H.  Keller, T.  Venkatesan, C.  W.  Chu, and R.  L.  
Greene, Phys.  Rev.  B {\bf 62}, R11 949 (2000).

\bibitem{Ban}A.  Banerjee, S.  Bahattachacharya, S.  Mollah, H.  
Sakata, H.  D.  Yang, B.  K.  Chaudhuri, Phys.  Rev.  B {\bf 68}, 186401 (2003).

\bibitem{Alex99} A. S. Alexandrov and P. E. Kornilovitch, Phys. Rev. 
Lett. \textbf{82}, 807 (1999).
\bibitem{Alex99C}A S Alexandrov and A M Bratkovsky, J. Phys.: Condens. 
Matter {\bf 11}, L531 (1999).

\bibitem{Austin}I. G. Austin and N. F. Mott, Adv. Phys. \textbf{18}, 
41 (1969).

\bibitem{Book}C.  Kittel, {\em Introduction to Solid State Physics}, 
7th ed., John Wiley and Sons, Inc., New York, 1996.

\bibitem{Emin} D. Emin and T. Holstein, Ann. Phys. (N.Y.), \textbf{53}, 
439 (1969).

\bibitem{Alexcond}A.  S.  Alexandrov and A.  M.  
Bratkovsky, Phys.  Rev.  Lett.  \textbf{82}, 141 (1999); A.  S.  Alexandrov and A.  M.  Bratkovsky, J.  
Phys.:Condens.  Matter, \textbf{11}, 1989 (1999).

\bibitem{Jung} J.  H.  Jung, K.  H.  Kim, and T.  W.  
Noh, E.  J.  Choi, and J.  J.  Yu, Phys.  Rev.  B {\bf 57}, R11 043 
(1998).

\bibitem{Kat}T.  Katsufuji, S.  Mori, M.  Masaki, Y.  Moritomo, N.  Yamamoto, 
and H.  Takagi, Phys.  Rev.  B {\bf 64}, 104419 (2001).

\bibitem{Pick1}W. E. Pickett and  D. J. Singh, Phys. 
Rev. B \textbf{55}, R8642 (1997).

\end{thebibliography}
\end{document}